\documentstyle[12pt,epsf]{article}
\newcommand{\be}{\begin{equation}}
\newcommand{\ee}{\end{equation}}
\newcommand{\bea}{\begin{eqnarray}}
\newcommand{\eea}{\end{eqnarray}}
\newcommand{\ep}{\varepsilon}

\newcommand{\nn}{\nonumber}

\begin{document}
 \thispagestyle{empty}
 \begin{flushright}
 {DESY 97-257} \\[5mm]
 {hep-ph/9801380} \\[5mm]
 {December 1997}           \\
\end{flushright}
 \vspace*{3cm}
 \begin{center}
 {\bf \Large
 Two-loop three-gluon vertex 
 in zero-momentum limit}
 \end{center}
 \vspace{1cm}
 \begin{center}
 A.~I.~Davydychev$^{a,b,}$\footnote{davyd@theory.npi.msu.su} , \ \
 P.~Osland$^{a,c,}$\footnote{Per.Osland@fi.uib.no} \ \
 \ \ and \ \
 O.~V.~Tarasov$^{d,}$\footnote{tarasov@ifh.de.
                                 On leave from
                                 Joint Institute for Nuclear Research,
                                 141980, Dubna, Russia.}\\
 \vspace{1cm}
$^{a}${\em
 Department of Physics, University of Bergen, \\
      All\'{e}gaten 55, N-5007 Bergen, Norway}
\\
\vspace{.3cm}
$^{b}${\em
 Institute for Nuclear Physics,
 Moscow State University, \\
 119899, Moscow, Russia}
\\
\vspace{.3cm}
$^c${\em 
 Deutsches Elektronen-Synchrotron DESY, 
 D-22603 Hamburg, Germany}
\\
\vspace{.3cm}
$^{d}${\em
 IfH, DESY-Zeuthen, Platanenallee 6, D-15738 Zeuthen, Germany}
\end{center}
 \hspace{3in}
 \begin{abstract}
The two-loop three-gluon vertex is calculated in an arbitrary covariant
gauge, in the limit when one of the external momenta vanishes.
The differential Ward--Slavnov--Taylor (WST) identity related to this
limit is discussed, and the relevant results for the ghost-gluon vertex
and two-point functions are obtained. Together with the
differential WST identity, they provide another independent way
for calculating the three-gluon vertex.
The renormalization of the results obtained is also presented.
 \end{abstract}

\newpage

\setcounter{footnote}{0}



\section{Introduction}
\setcounter{equation}{0}

Jet studies are becoming increasingly precise, both as a 
testing ground for QCD, and as a background
for new physics (e.g.\ Higgs searches).
Increasing precision, among other things, requires knowledge
of the fundamental QCD vertices to higher loops.

The one-loop vertices have been known for quite some time.
Celmaster and Gonsalves presented in 1979 \cite{CG}
the one-loop result for the three-gluon vertex, for off-shell gluons,
restricted to the symmetric case, $p_1^2=p_2^2=p_3^3$,
in an arbitrary covariant gauge. The result of \cite{CG}
was confirmed by Pascual and Tarrach \cite{PT}.
Ball and Chiu then in 1980 considered the general off-shell case,
but restricted to the Feynman gauge \cite{BC2}.
Later, various on-shell results have also been given,
by Brandt and Frenkel \cite{BF},
restricted to the infrared-singular parts only
(in an arbitrary covariant gauge),
and by Nowak, Prasza{\l}owicz and S{\l}omi{\'n}ski 
\cite{NPS}, who also gave the finite parts for the case of
two gluons being on-shell (in Feynman gauge).
The most general results, valid for arbitrary values
of the space-time dimension and the covariant-gauge 
parameter, have been presented in our previous paper \cite{DOT1}.
Some results for the one-loop quark-gluon vertex (or its 
Abelian part which is related to the QED vertex) can be
found in \cite{qqg}.

The present paper is devoted to a study of two-loop
corrections to the three-gluon vertex
in the zero-momentum limit.
This limit refers to the case when one gluon     
has vanishing momentum.
The remaining two momenta must then   
be equal and opposite, so there is only one dimensionful
scale, $p^2$. 
In this limit, the renormalized expressions for QCD vertices 
in the Feynman gauge have been presented by Braaten and
Leveille \cite{BL}.
Information about Green functions is also required
for calculation of certain quantities related to the
renormalization group equations, such as the $\beta$ function
and anomalous dimensions. The two-loop-order contributions 
to these quantities were calculated
in refs.~\cite{Caswell,Jones,VlaTar,EgTar}, whereas the 
three-loop-order results were obtained in \cite{TVZh,LV}. 
Moreover, recently the four-loop-order expressions became
available \cite{LvRV}.

When massless quarks are considered, the scalar 
functions corresponding to the coefficients of different
tensor structures are in the zero-momentum limit rather simple: 
apart from non-trivial coefficients, they are given by $p^2$ raised 
to some power (determined by the dimension of space-time).
Also, the tensorial structure is considerably simpler than 
in the general case.
Although the zero-momentum limit has limited physical
applications,
it serves as an important reference point, against which more
general results can be checked.

With one gluon momentum vanishing, there are two Ward-Slavnov-Taylor
(WST) identities, one corresponding to the vanishing momentum, and one
corresponding to the finite momentum.
The identity corresponding to the vanishing momentum turns out to be
a differential identity. In this case, the three-gluon vertex
can actually be completely constructed from the two-point functions
and the ghost-gluon vertex, with no additional transverse term.

In the present paper, we realize two ways to calculate
the two-loop three-gluon vertex in an arbitrary covariant gauge. 
One of them is  
a straightforward calculation of all diagrams contributing to
the three-gluon vertex at this order. Another way is based
on using the results for the ghost-gluon vertex and the
two-point functions, together with the corresponding WST   
identities. The renormalized expressions are also obtained. 

\section{Preliminaries}
\setcounter{equation}{0}

The lowest-order gluon propagator is
\begin{equation}
\label{gl_prop}
\delta^{a_1 a_2} \; \frac{1}{p^2}
\left( g_{\mu_1 \mu_2} - \xi \; \frac{p_{\mu_1} p_{\mu_2}}{p^2}
\right) ,
\end{equation}
where $\xi\equiv1-\alpha$ is the gauge parameter corresponding 
to a general covariant gauge, 
defined such that $\xi=0$ ($\alpha=1$) is the Feynman gauge.
Here and henceforth, a causal prescription is understood,
$1/p^2 \rightarrow 1/(p^2 +\mbox{i}0)$.

The three-gluon vertex is defined as
\begin{equation}
\label{ggg}
\Gamma_{\mu_1 \mu_2 \mu_3}^{a_1 a_2 a_3}(p_1, p_2, p_3)
\equiv  - \mbox{i} \; g \;
f^{a_1 a_2 a_3} \; \Gamma_{\mu_1 \mu_2 \mu_3}(p_1, p_2, p_3) ,
\end{equation}
where $f^{a_1 a_2 a_3}$
are the totally antisymmetric colour structures corresponding
to the adjoint representation of the gauge group
(for example, $\mbox{SU}(N)$ or any other semi-simple
gauge group).
In fact, also completely symmetric
colour structures $d^{a_1 a_2 a_3}$ might be considered,
but they do not appear in the perturbative calculation of
QCD three-point vertices at the one- and two-loop level.
Since the gluons are bosons, and since the colour structures
$f^{a_1 a_2 a_3}$ are antisymmetric,
$\Gamma_{\mu_1 \mu_2 \mu_3}(p_1, p_2, p_3)$ must also be 
{\em antisymmetric}
under any interchange of a pair of gluon momenta and the 
corresponding Lorentz indices.

When one of the momenta is zero, the three-gluon vertex
contains only two tensor structures\footnote{This is a corollary
of the differential WST identity, see in section~3.},
\be
\label{BL-decomp}
\Gamma_{\mu_1 \mu_2 \mu_3}(p, -p, 0)
= \left( 2 g_{\mu_1 \mu_2} p_{\mu_3} - g_{\mu_1 \mu_3} p_{\mu_2}
         - g_{\mu_2 \mu_3} p_{\mu_1} \right)  T_1(p^2)
- p_{\mu_3} \left( g_{\mu_1 \mu_2}
                      - \frac{p_{\mu_1} p_{\mu_2}}{p^2} \right) 
                                               T_2(p^2) .
\ee
In this decomposition, we basically adopt the notation of \cite{BL} 
for the scalar functions $T_i(p^2)$.
The first tensor structure on the r.h.s. of eq.~(\ref{BL-decomp})
corresponds to the lowest-order vertex.  
There is the following correspondence between the
functions $T_i$ and the scalar 
functions $A$ and $C$ used in \cite{BC2} (cf. also in \cite{DOT1}):
\begin{equation}
T_1(p^2)  \leftrightarrow A(p^2, p^2; 0) , \hspace{7mm}
T_2(p^2)  \leftrightarrow -2 p^2 C(p^2, p^2; 0) .
\end{equation}
At the lowest, ``zero-loop'' order, the Yang--Mills 
term of the QCD Lagrangian yields\footnote{We include the
contribution $T_1^{(0)}=1$ into the definition of $T_1(p^2)$,
eq.~(\ref{BL-decomp}).}
\be
\label{T_i^0}
T_1^{(0)}=1, \hspace{10mm} T_2^{(0)}=0 .
\ee

For a quantity $X$ (e.g.\ any of the scalar functions contributing 
to the propagators or the vertices), we shall denote the 
zero-loop-order contribution as $X^{(0)}$ (cf. eq.~(\ref{T_i^0})), 
the one-loop-order contribution as
$X^{(1)}$, and the two-loop-order contribution as $X^{(2)}$.
In this paper, as a rule,
\begin{equation}
X^{(L)}=X^{(L,\xi)} + X^{(L,q)}, 
\end{equation}
where $X^{(L,\xi)}$ denotes the contribution 
of gluon and ghost
loops in a general covariant gauge (\ref{gl_prop}) (in particular,
$X^{(L,0)}$ corresponds to the Feynman gauge, $\xi=0$),
while $X^{(L,q)}$ represents the contribution of the
quark loops.

The ghost-gluon vertex can be represented as
\begin{equation}
\label{ghg}
\widetilde{\Gamma}_{\mu_3}^{a_1 a_2 a_3}(p_1, p_2; p_3)
\equiv -\mbox{i} g \; f^{a_1 a_2 a_3} \;
{p_1}^{\mu} \; \widetilde{\Gamma}_{\mu \mu_3}(p_1, p_2; p_3) ,
\end{equation}
where $p_1$ is the out-ghost momentum, $p_2$ is the in-ghost momentum,
$p_3$ and $\mu_3$ are the momentum and the Lorentz index of the gluon
(all momenta are ingoing). For $\widetilde{\Gamma}_{\mu \mu_3}$, 
the following decomposition
was used in \cite{BC2}:
\begin{eqnarray}
\label{BC-ghg}
\widetilde{\Gamma}_{\mu \mu_3}(p_1,p_2;p_3)
= g_{\mu \mu_3} a(p_3,p_2,p_1)
- {p_3}_{\mu} {p_2}_{\mu_3} b(p_3,p_2,p_1)
+ {p_1}_{\mu} {p_3}_{\mu_3} c(p_3,p_2,p_1)
\nonumber \\
+ {p_3}_{\mu} {p_1}_{\mu_3} d(p_3,p_2,p_1)
+ {p_1}_{\mu} {p_1}_{\mu_3} e(p_3,p_2,p_1) .
\end{eqnarray}
At the ``zero-loop'' level,
\begin{equation}
\label{ghg0}
\widetilde{\Gamma}^{(0)}_{\mu \mu_3} = g_{\mu \mu_3} ,
\end{equation}
and therefore all the scalar functions involved in  (\ref{BC-ghg})
vanish at this order, except one, $a^{(0)}=1$.

We shall need the results for the ghost-gluon vertex (\ref{BC-ghg})
for two different configurations: 
(i) when the gluon momentum, $p_3$,
is zero and (ii) when the in-ghost momentum, $p_2$, is zero.
In the former case, we get
\be
\label{ghg1}
\widetilde{\Gamma}_{\mu \mu_3}(-p,p;0)
= g_{\mu \mu_3} a_3(p^2)
+ p_{\mu} p_{\mu_3} e_3(p^2) , \hspace{4mm}
a_3(p^2)\equiv a(0,p,-p), \hspace{4mm} e_3(p^2)\equiv e(0,p,-p),
\ee
whereas in the latter case we obtain
\be
\label{ghg2}
\widetilde{\Gamma}_{\mu \mu_3}(p,0;-p)
= g_{\mu \mu_3} a_2(p^2)
+ p_{\mu} p_{\mu_3} e'_2(p^2) , \hspace{4mm}
a_2(p^2)\equiv a(-p,0,p), \hspace{4mm} e'_2(p^2)\equiv e'(-p,0,p),
\ee 
with
\be
e'(p_3,p_2,p_1) \equiv 
e(p_3,p_2,p_1) - c(p_3,p_2,p_1) - d(p_3,p_2,p_1)  .
\ee
We shall also denote
\be
\label{d_2}
d_2(p^2)\equiv d(-p,0,p).
\ee
We do not need to
consider $\widetilde{\Gamma}_{\mu\mu_3}(0,p,-p)$ ($p_1=0$)
because it does not enter the WST identities (see in section~3).
Moreover, the proper ghost-gluon vertex (\ref{ghg})
vanishes in this limit, for it contains $p_1^{\;\;\mu}$.

The gluon polarization operator is defined as
\begin{equation}
\label{gl_po}
\Pi_{\mu_1 \mu_2}^{a_1 a_2}(p)
\equiv - \delta^{a_1 a_2}
\left( p^2 g_{\mu_1 \mu_2} - {p}_{\mu_1}{p}_{\mu_2} \right) J(p^2),
\end{equation}
while the ghost self energy is\footnote{There was a misprint in
eq.~(2.8) of \cite{DOT1}: 
$G(p^2)$ should read $\left[G(p^2)\right]^{-1}$.}
\begin{equation}
\label{gh_se}
\widetilde{\Pi}^{a_1 a_2}(p^2) = \delta^{a_1 a_2} \; p^2 \;
\left[G(p^2)\right]^{-1} .
\end{equation}
In the lowest-order approximation $J^{(0)}=G^{(0)}=1$.

\section{WST identity in the zero-momentum limit}
\setcounter{equation}{0}

In a covariant gauge, the Ward--Slavnov--Taylor (WST) identity 
\cite{WST}
for the three-gluon vertex is of the following form (see e.g.\ in
\cite{MarPag}):
\begin{eqnarray}
\label{WST}
p_3^{\mu_3} \; \Gamma_{\mu_1 \mu_2 \mu_3}(p_1, p_2, p_3)
= - J(p_1^2) \; G(p_3^2) \;
\left( g_{\mu_1 \;\;\;}^{\;\;\; \mu_3}\; p_1^2
       - {p_1}_{\mu_1} \; {p_1}^{\mu_3} \right) \;
\widetilde{\Gamma}_{\mu_3 \mu_2}(p_1, p_3; p_2)
\hspace{1mm}
\nonumber \\
 + J(p_2^2) \; G(p_3^2) \;
\left( g_{\mu_2 \;\;\;}^{\;\;\; \mu_3}\; p_2^2
       - {p_2}_{\mu_2} \; {p_2}^{\mu_3} \right) \;
\widetilde{\Gamma}_{\mu_3 \mu_1}(p_2, p_3; p_1) .
\end{eqnarray}
It is easy to see that the $c$ and $e$ functions
from the ghost-gluon vertex (\ref{BC-ghg}) do not contribute
to this identity.

Consider what follows from (\ref{WST}) in the limit when one of the
momenta vanishes.
We should distinguish between two different cases: when the vanishing
momentum is the one with which the three-gluon vertex is contracted, 
and when it is not.
In the former case, we obtain a differential identity, whereas in
the latter case we get an ordinary identity.

In the differential case, we should consider 
$p_3\equiv \delta \to 0, \; p_1\equiv p, \; p_2=-p-\delta$. 
We do not need the terms of
order $\delta^2$ and higher. In particular,
$G(\delta^2)=G(0)+{\cal{O}}(\delta^2)$ and, for {\em massless}
quarks, $G(0)=1$.
When we expand the r.h.s.\ of eq.~(\ref{WST}) in $\delta$, the
lowest (``constant'') term disappears, so only the term linear in
$\delta$ is relevant. Differentiating both
sides with respect to $\delta^{\mu_3}$ and putting $\delta=0$,
we get
\begin{eqnarray}
\label{WST-diff}
\Gamma_{\mu_1 \mu_2 \mu_3}(p, -p, 0)
= \left( 2 g_{\mu_1 \mu_2} p_{\mu_3}  - g_{\mu_1 \mu_3} p_{\mu_2}  
      - g_{\mu_2 \mu_3} p_{\mu_1} \right) \;
\left[ a_2(p^2) - p^2 d_2(p^2) \right] \; J(p^2) \; G(0)
\hspace{2.5mm}
\nonumber \\
+ 2 p_{\mu_3} 
\left(\! g_{\mu_1 \mu_2} \!-\! \frac{p_{\mu_1} p_{\mu_2}}{p^2}\!\right)\!
\left[ \left( p^2 d_2(p^2) \!+\! \widetilde{a}_2(p^2)
\!-\!p^2\frac{\mbox{d} a_2(p^2)}{\mbox{d} p^2} \right) J(p^2) 
\!+\! p^2 a_2(p^2) \frac{\mbox{d} J(p^2)}{\mbox{d} p^2} \right]G(0) ,
\nn \\
{\hspace{10mm}}
\end{eqnarray}
where the functions $a_2(p^2)$ and $d_2(p^2)$ are defined
in eqs.~(\ref{ghg2}) and (\ref{d_2}), respectively.
The function $\widetilde{a}_2(p^2)$ is defined as
\begin{equation}
\label{a_tilde_2}
\widetilde{a}_2(p^2) \equiv 
\left. {p_1}_{\sigma}\; \frac{\partial}{\partial {p_1}_{\sigma}} \;
          a(p_3,-p_1-p_3,p_1)\right|_{p_1=-p_3=p} \; \; .
\end{equation}
It can be calculated directly at the diagrammatic level (see in
section~5). 

Considering contraction with a non-zero momentum, we get from
eq.~(\ref{WST})
\begin{equation}
\label{WST-ord}
p^{\mu_1} \Gamma_{\mu_1 \mu_2 \mu_3}(p, -p, 0)
= - J(p^2) G(p^2) a_3(p^2)
\left( g_{\mu_2 \mu_3} p^2 - p_{\mu_2} p_{\mu_3} \right) ,
\end{equation}
where $a_3(p^2)$ is defined in eq.~(\ref{ghg1}).
Contracting eq.~(\ref{WST-diff}) with $p^{\mu_1}$
we get a different representation which should be equal to the r.h.s.\
of eq.~(\ref{WST-ord}). Therefore, the following relation should hold:
\begin{equation}
\label{adG}
G(0) \left[ a_2(p^2) - p^2 d_2(p^2) \right] = G(p^2) a_3(p^2) .
\end{equation}

Using eq.~(\ref{adG}), the differential WST identity (\ref{WST-diff})
can be re-written in a way
which involves just the $a$ functions from the ghost-gluon vertex:
\begin{eqnarray}
\label{WST-diff3}
\Gamma_{\mu_1 \mu_2 \mu_3}(p, -p, 0)
= - \left[
p_{\mu_1}
\left( g_{\mu_2 \mu_3} \!-\! \frac{p_{\mu_2} p_{\mu_3}}{p^2} \right)
+ p_{\mu_2}
\left( g_{\mu_1 \mu_3} \!-\! \frac{p_{\mu_1} p_{\mu_3}}{p^2} \right)
\right] 
a_3(p^2) G(p^2) J(p^2)
\nonumber \\
+ 2 p_{\mu_3}
\left( g_{\mu_1 \mu_2} \!-\! \frac{p_{\mu_1} p_{\mu_2}}{p^2} \right)
G(0) 
\left[
a_2(p^2) \frac{\mbox{d}}{\mbox{d} p^2}\left( p^2 J(p^2) \right)
\!-\! p^2 J(p^2) \frac{\mbox{d}a_2(p^2)}{\mbox{d} p^2}
\!+\! \widetilde{a}_2(p^2) J(p^2)
\right] .
\hspace{4mm}
\end{eqnarray}

For the scalar functions $T_i(p^2)$, the WST identity gives
\be
\label{T1-WST}
T_1(p^2)=a_3(p^2) \; G(p^2) \; J(p^2) ,
\ee
\be
\label{T2-WST}
T_2(p^2)= 2 T_1(p^2) 
- 2 G(0) 
\left[ 
a_2(p^2) \frac{\mbox{d}}{\mbox{d} p^2}\left( p^2 J(p^2) \right)
- p^2 J(p^2) \frac{\mbox{d}a_2(p^2)}{\mbox{d} p^2}
+ \widetilde{a}_2(p^2) J(p^2)
\right] .
\ee

Therefore, the differential WST identity  makes it possible to 
define the whole
three-gluon vertex (not only its longitudinal part)
in terms of two-point functions and the ghost-gluon vertex.
Moreover, it can be used as another independent way, in addition to
the direct calculation, to obtain results for the three-gluon
vertex.

\section{Results for the three-gluon vertex}
\setcounter{equation}{0}

We shall use dimensional regularization \cite{dimreg},
with the space-time dimension $n=4-2\ep$.
The results for unrenormalized one-loop contributions 
to the scalar functions 
$T_1(p^2)$ and $T_2(p^2)$ (in arbitrary space-time dimension)
can be found in ref.~\cite{DOT1}, eqs.~(4.30), (4.31), (4.33)
and (4.34). Expanding them in $\ep$ we get\footnote{In 
all unrenormalized expressions given in sections~4--7 
and in Appendix~A, the {\em bare} quantities $g^2=g_B^2$
and $\xi=\xi_B$ are understood, i.e. the same as those given
in the lowest-order functions (\ref{gl_prop})--(\ref{ggg}). 
When the renormalization
is discussed, these bare quantities get a subscript ``$B$''
(see in section~8).}
\bea
\label{T1xi}
T_1^{(1,\xi)}(p^2)
=C_A  \frac{g^2 \; \eta}{(4\pi)^{n/2}} \; (-p^2)^{-\ep}
\left\{ 
\frac{1}{\ep} \left( -\frac{2}{3} - \frac{3}{4}\xi \right)
-\frac{35}{18}+\frac{1}{2}\xi-\frac{1}{4}\xi^2
\right.
\nn \\
\left.
+\ep \left(-\frac{107}{27}+\xi-\frac{1}{2}\xi^2 \right) \right\}
+ {\cal{O}}(\ep^2),
\eea
\be
\label{T1q}
T_1^{(1,q)}(p^2)=
T  \frac{g^2 \; \eta}{(4\pi)^{n/2}} \; (-p^2)^{-\ep}
\left\{ \frac{4}{3\ep}+\frac{20}{9}+\frac{112}{27}\ep\right\} 
+ {\cal{O}}(\ep^2),
\ee
\be
\label{T2xi}
T_2^{(1,\xi)}(p^2)=
C_A  \frac{g^2 \; \eta}{(4\pi)^{n/2}} \; (-p^2)^{-\ep}
\left\{ -\frac{4}{3}-2\xi+\frac{1}{4}\xi^2
        +\ep\left( -\frac{26}{9}-\xi+\frac{1}{4}\xi^2 \right)
\right\}
+ {\cal{O}}(\ep^2),
\ee
\be
\label{T2q}
T_2^{(1,q)}(p^2)=
T  \frac{g^2 \; \eta}{(4\pi)^{n/2}} \; (-p^2)^{-\ep}
\left\{\frac{8}{3}+\frac{40}{9}\ep\right\}
+ {\cal{O}}(\ep^2).
\ee
In these equations,
we use the standard notation $C_A$ for the eigenvalue
of the quadratic Casimir operator in the adjoint representation,
\begin{equation}
\label{C_A}
f^{acd}f^{bcd} = C_A \, \delta^{ab} \hspace{5mm}
(C_A = N \; \mbox{for the SU($N$) group}) .
\end{equation}
Furthermore,
\begin{equation}
\label{T_R}
T\equiv N_f T_R, \hspace{7mm}
T_R = {\textstyle{1\over8}} \; \mbox{Tr}(I) = {\textstyle{1\over2}} \; ,
\end{equation}
where $I$ is the ``unity'' in the space of Dirac matrices
(we assume that $\mbox{Tr}(I)=4$),
$N_f$ is the number of quarks and
\begin{equation}
\label{eta}    
\eta \equiv
\frac{\Gamma^2(\frac{n}{2}-1)}{\Gamma(n-3)} \;  
     \Gamma(3-{\textstyle{n\over2}}) =
\frac{\Gamma^2(1-\varepsilon)}{\Gamma(1-2\varepsilon)} \;
\Gamma(1+\varepsilon) 
=e^{-\gamma\ep}\left(1-\frac{1}{12}\pi^2\ep^2 + {\cal{O}}(\ep^3)
\right).
\end{equation}
Here $\gamma\simeq 0.57721566...$ is the Euler constant.
The $\ep$ terms in the expressions (\ref{T1xi})--(\ref{T2q})
are needed when these expressions are multiplied by terms which
diverge like $1/\ep$, e.g.,
for the calculation of reducible unrenormalized 
two-loop-order contributions. 
The $\ep$ terms are also necessary for getting the
renormalized two-loop-order results, see section~8.

The diagrams contributing to the three-gluon vertex at the two-loop
level are shown in Fig.~1\footnote{To produce the figures,
the {\sf AXODRAW} package \cite{axodraw} was used.}.
Each diagram should be considered with two other ``rotations'',
corresponding to permutations of the external legs.
The grey blob corresponds to a sum of all one-loop 
contributions to the gluon polarization operator,
including the gluon, ghost and quark loops 
insertions\footnote{Here and
henceforth, we do not show contributions involving tadpole-like 
insertions
which vanish in the framework of dimensional regularization
\cite{dimreg}.}, cf. Fig.~2a of \cite{DOT1}.
Note that non-planar graphs do not contribute to the two-loop
vertex, since their over-all colour factors vanish, due to the
Jacobi identity (cf. Fig.~6 of ref.~\cite{Cvit} where this
is explained). 

When one external momentum vanishes, technically the problem
reduces to the calculation of two-point two-loop Feynman 
integrals. To calculate the occurring integrals with higher
powers of the propagators, the integration-by-parts procedure
\cite{ibp} has been used. For the integrals with numerators,
some other known algorithms \cite{ibp} (see also in \cite{PLB'91})
were employed. 
Straightforward calculation of the sum 
of all these contributions\footnote{For this calculation, two 
independent computer programs written in {\sf REDUCE} \cite{reduce}
and {\sf FORM} \cite{form} were used.}  
yields the following results for 
the unrenormalized scalar functions:
\bea
\label{T12xi}
T_1^{(2,\xi)}(p^2) 
= C_A^2 \frac{g^4 \; \eta^2}{(4\pi)^n}  (-p^2)^{-2\ep}
\left\{
\frac{1}{\ep^2}\left( 
- \frac{13}{8} \!-\! \frac{7}{16} \xi \!+\! \frac{15}{32} \xi^2 \right)
\!+\! \frac{1}{\ep}
\left( - \frac{311}{48}\!+\! \frac{13}{96} \xi \!-\! \frac{29}{48} \xi^2
\!+\! \frac{7}{16} \xi^3 \right)
\hspace*{-8mm}
\right.
\nn \\
\left.
- \frac{6965}{288}- \frac{1}{4} \zeta_3 - \frac{509}{576} \xi
+ \frac{15}{8} \xi \zeta_3  - \frac{115}{144} \xi^2
+ \frac{13}{16} \xi^3 + \frac{1}{16} \xi^4 
\right\} 
+ {\cal{O}}(\ep),
\hspace{4mm}
\eea
\bea
\label{T12q}
T_1^{(2,q)}(p^2) 
= C_A T \frac{g^4 \; \eta^2}{(4\pi)^n} (-p^2)^{-2\ep}
\left\{
\frac{1}{\ep^2}\left(  \frac{5}{2}  - \xi \right)
+\frac{1}{\ep}\left(  \frac{97}{12}- \frac{1}{3} \xi- \frac{2}{3} \xi^2 
\right)
\right.
\hspace{10mm}
\nn \\
\left.
+ \frac{1675}{72}+ 8 \zeta_3+ \frac{16}{9} \xi- \frac{22}{9} \xi^2
\right\}
\nn \\
+ C_F T \frac{g^4 \; \eta^2}{(4\pi)^n} \; (-p^2)^{-2\ep}
\left\{ 
\frac{2}{\ep}+ \frac{55}{3} - 16 \zeta_3 \right\}
+ {\cal{O}}(\ep),
\eea
\bea
\label{T22xi}
T_2^{(2,\xi)}(p^2) 
= C_A^2 \frac{g^4 \; \eta^2}{(4\pi)^n} \; (-p^2)^{-2\ep}
\left\{
\frac{1}{\ep} \left(
-\frac{22}{3} - \frac{11}{6} \xi + \frac{8}{3} \xi^2
-\frac{7}{16} \xi^3 \right)
\hspace{27mm}
\right.
\nn \\
\left.
 - \frac{1013}{36} - \zeta_3 + \frac{13}{9} \xi 
 - \frac{1}{2} \xi \zeta_3 
 - \frac{83}{144} \xi^2 + \frac{3}{4} \xi^3 - \frac{1}{8} \xi^4
\right\}
+ {\cal{O}}(\ep),
\eea
\bea
\label{T22q}
T_2^{(2,q)}(p^2)  
= C_A T \frac{g^4 \; \eta^2}{(4\pi)^n} \; (-p^2)^{-2\ep}
\left\{
\frac{1}{\ep} \left( \frac{32}{3}  - \frac{16}{3} \xi 
+ \frac{2}{3} \xi^2 \right)
 + \frac{289}{9}- \frac{133}{18} \xi+ \frac{4}{9} \xi^2 
\right\}
\nn \\
+ 8 C_F T \frac{g^4 \; \eta^2}{(4\pi)^n} \; (-p^2)^{-2\ep}
+ {\cal{O}}(\ep),
\hspace{20mm}
\eea
where 
$\zeta_3\equiv\zeta(3)
=\sum_{j=1}^{\infty} j^{-3} \simeq       
1.2020569...$
is the value of Riemann's zeta function;
$C_F$ is the eigenvalue of the quadratic
Casimir operator
in the fundamental representation. For the $\mbox{SU}(N)$ group,
$C_F=(N^2-1)/(2N)$.

\section{Results for the ghost-gluon vertex}
\setcounter{equation}{0}

In order to check the WST identity, we need results for the
ghost-gluon vertex in two limits corresponding to eqs.~(\ref{ghg1})
and (\ref{ghg2}). We shall also need the derivative 
$\widetilde{a}_2(p^2)$, eq.~(\ref{a_tilde_2}). 

The relevant one-loop results (for an arbitrary $n$)
are listed in Appendix~A. Expanding them in $\ep$ we get 
\be
a_3^{(1)}(p^2) =
C_A \; \frac{g^2\; \eta}{(4\pi)^{n/2}} \; (-p^2)^{-\ep}
\; (1-\xi)
\left\{ \frac{1}{2\ep} + \frac{1}{2} + \ep \right\} 
+ {\cal{O}}(\ep^2),
\ee   
\be
a_2^{(1)}(p^2) =
C_A \; \frac{g^2\; \eta}{(4\pi)^{n/2}} \; (-p^2)^{-\ep}
\; (1-\xi)
\left\{  \frac{1}{2\ep} + \frac{1}{4} \xi + \frac{1}{2} \xi \ep 
\right\} + {\cal{O}}(\ep^2),
\ee
\be
\widetilde{a}_2^{(1)}(p^2)
= C_A \;\frac{g^2\; \eta}{(4\pi)^{n/2}}  \; (-p^2)^{-\ep}
\left\{
\frac{1}{\ep}\left(\frac{1}{2}+\frac{1}{4}\xi\right)
+\frac{1}{4}\xi+\frac{1}{8}\xi^2
+\ep\left(1-\frac{1}{4}\xi+\frac{3}{8}\xi^2\right) \right\}
+ {\cal{O}}(\ep^2),
\ee
\be
p^2 e_3^{(1)}(p^2)
= C_A \; \frac{g^2\; \eta}{(4\pi)^{n/2}}  \; (-p^2)^{-\ep}
\left\{ \frac{1}{2} + \frac{1}{4}\xi+\ep\right\}
+ {\cal{O}}(\ep^2) ,
\ee
\be
p^2 {e'}_2^{(1)}(p^2)  
= C_A \; \frac{g^2\; \eta}{(4\pi)^{n/2}}  \; (-p^2)^{-\ep}
(1-\xi) (2-\xi) 
\left\{ \frac{1}{4} + \frac{1}{2}\ep \right\}
+ {\cal{O}}(\ep^2) .
\ee

Two-loop contributions to the ghost-gluon vertex are shown in Fig.~2.
As in the case of the three-gluon vertex (cf. Fig.~1), non-planar
graphs do not contribute (cf. ref.~\cite{Cvit}).
Straightforward calculation gives the following results:
\bea
a_3^{(2,\xi)}(p^2) =
C_A^2 \frac{g^4 \; \eta^2}{(4\pi)^n} \; (-p^2)^{-2\ep}
\left\{ 
\frac{1}{\ep^2} \left(  \frac{5}{8}- \frac{7}{8}\xi
+ \frac{1}{4} \xi^2 \right)
+\frac{1}{\ep}  \left( \frac{13}{8} - \frac{35}{16} \xi
+ \frac{9}{16} \xi^2 \right)
\right.
\hspace{8mm}
\nn \\
\left.
+ \frac{257}{48} - \frac{1}{2} \zeta_3 - \frac{635}{96}\xi
- \frac{1}{8} \xi \zeta_3 + \frac{23}{16} \xi^2
+ \frac{3}{16} \xi^2 \zeta_3
\right\}+ {\cal{O}}(\ep),
\eea
\be
a_3^{(2,q)}(p^2) =
 \frac{1}{4}
C_A T \frac{g^4 \; \eta^2}{(4\pi)^n} \; (-p^2)^{-2\ep}
\; + {\cal{O}}(\ep), 
\ee
\be
p^2 e_3^{(2,\xi)}(p^2) =\!
C_A^2 \frac{g^4 \; \eta^2}{(4\pi)^n} (-p^2)^{-2\ep}
\left\{
\frac{1}{\ep}
\left(  \frac{5}{2} \!+\!\frac{1}{2} \xi \!-\! \frac{1}{4}\xi^2\right)
\!+\! \frac{65}{6}\!+\! \frac{1}{8} \zeta_3 \!-\!\frac{11}{12}\xi
\!+\! \frac{5}{16} \xi \zeta_3 \!-\! \frac{3}{16}\xi^2 
\right\}
\!+\! {\cal{O}}(\ep),
\ee
\be
p^2 e_3^{(2,q)}(p^2) =
C_A T \frac{g^4 \; \eta^2}{(4\pi)^n} \; (-p^2)^{-2\ep}
\left\{ - \frac{1}{\ep} - 4 \right\}
+ {\cal{O}}(\ep),
\ee
\bea
a_2^{(2,\xi)}(p^2) =
C_A^2 \frac{g^4 \; \eta^2}{(4\pi)^n} \; (-p^2)^{-2\ep}
(1-\xi)
\left\{ 
\frac{1}{\ep^2}\left(\frac{5}{8}  -\frac{1}{4} \xi \right)
+\frac{1}{\ep}\left( \frac{19}{24} + \frac{13}{48}\xi
 - \frac{3}{8} \xi^2 \right)
\right.
\hspace{8mm}
\nn \\
\left.
+ \frac{227}{72} - \zeta_3+ \frac{53}{144}\xi- \frac{13}{16}\xi^2
-\frac{1}{16} \xi^3 
\right\}+ {\cal{O}}(\ep),
\eea
\be
a_2^{(2,q)}(p^2) =
C_A T \frac{g^4 \; \eta^2}{(4\pi)^n} \; (-p^2)^{-2\ep}
\; (1-\xi)^2 
\left\{    - \frac{1}{3\ep} - \frac{11}{9}  \right\}
+ {\cal{O}}(\ep),
\ee
\bea
p^2 {e'}_2^{(2,\xi)}(p^2) =
C_A^2 \frac{g^4 \; \eta^2}{(4\pi)^n} \; (-p^2)^{-2\ep}
(1-\xi)
\left\{
\frac{1}{\ep}
\left(\frac{5}{6} - \frac{5}{6}\xi +  \frac{3}{8}\xi^2 \right)
\right.
\hspace{28mm}
\nn \\
\left.
+ \frac{89}{36} + \frac{5}{8} \zeta_3- \frac{65}{36}\xi
 - \frac{3}{16} \xi\zeta_3+ \frac{13}{16}\xi^2
+\frac{1}{16} \xi^3 
\right\}
+ {\cal{O}}(\ep),
\eea
\be
p^2 {e'}_2^{(2,q)}(p^2) =
C_A T \frac{g^4 \; \eta^2}{(4\pi)^n} \; (-p^2)^{-2\ep}
\; (1-\xi)^2
\left\{  \frac{1}{3\ep} + \frac{11}{9} \right\}
+ {\cal{O}}(\ep).
\ee   

The derivative (\ref{a_tilde_2}) has been
calculated in the following way. The momenta $p_1$ and $p_3$
are considered as independent variables, whereas
$p_2=-p_1-p_3$. Therefore, the momentum $p_1$ flows from
the in-ghost leg to the out-ghost leg. An unambiguous
$p_1$ path inside the diagram can be chosen as the one
coinciding with the ghost line. This is convenient,
since all we need to differentiate are just two types of objects: 
ghost propagators and ghost-gluon vertices occurring along this path.
In this way, we avoid differentiating gluon propagators
and three-gluon vertices. We also avoid getting third powers
of propagators. 

Technically, this was realized as follows.
The list of diagrams
contributing to the ghost-gluon vertex, Fig.~2, was taken. Then,
the propagators and vertices along the ghost path were
``marked'' by introducing an extra argument (say, $z$).
Of course, the closed ghost loops should not be marked.
Then, the derivative with respect to $z$ was considered,
and the rules for differentiating the ghost-gluon vertex
and the ghost propagator (with subsequent contraction
with ${p_1}_{\mu_1}$) were supplied.
It is very important that we do not really need expressions
with different momenta; we just formally differentiate along the
ghost line, and then perform all calculations for $p_1=-p_3=p$,
$p_2=0$. Finally, extracting the coefficient of $g_{\mu \mu_3}$ 
gives the following results for the function (\ref{a_tilde_2}):
\bea
\label{a2z_p2_xi}
\widetilde{a}_2^{(2,\xi)}(p^2)
= C_A^2 \frac{g^4 \; \eta^2}{(4\pi)^n} \; (-p^2)^{-2\ep}
\left\{ 
\frac{1}{\ep^2}
\left( \frac{3}{2}+\frac{5}{16}\xi- \frac{5}{32}\xi^2 \right)
+\frac{1}{\ep}
\left( \frac{121}{48} + \frac{185}{96}\xi + \frac{1}{24}\xi^2
       - \frac{7}{32}\xi^3 \right)
\right.
\hspace*{-9mm}
\nn \\
\left.
+ \frac{3085}{288}+ \frac{1}{4} \zeta_3+ \frac{1265}{576}\xi
- \frac{7}{8} \xi\zeta_3 + \frac{389}{288}\xi^2
- \frac{13}{16}\xi^3 - \frac{1}{32} \xi^4
\right\} + {\cal{O}}(\ep), 
\hspace{5mm}
\eea
\bea
\label{a2z_p2_q}
\widetilde{a}_2^{(2,q)}(p^2)
= C_A T \frac{g^4 \; \eta^2}{(4\pi)^n} \; (-p^2)^{-2\ep}
\left\{  
-\frac{1}{2\ep^2}
+\frac{1}{\ep}
\left( - \frac{17}{12} - \frac{2}{3}\xi +\frac{1}{6} \xi^2 \right)
\right.
\nn \\
\left.
- \frac{239}{72} - \frac{79}{36}\xi + \frac{7}{9}\xi^2 
\right\} 
+ {\cal{O}}(\ep). 
\eea

\section{Results for the two-point functions}
\setcounter{equation}{0}

Before presenting the results, let us make some general remarks.
According to eq.~(\ref{gl_po}), the gluon polarization operator is 
proportional to
\be    
J(p^2) = 1 + J^{(1)}(p^2) + J^{(2)}(p^2) + \ldots
\ee
Two-loop contributions to the gluon polarization operator
are shown in Fig.~3.
The gluon propagator is proportional to
\be
\frac{1}{J(p^2)}
\left( g_{\mu_1 \mu_2} - \frac{p_{\mu_1} p_{\mu_2}}{p^2} \right)
+ (1-\xi) \frac{p_{\mu_1} p_{\mu_2}}{p^2} .
\ee
Therefore, the transverse part of the propagator
is proportional to
\be   
\left[ J(p^2) \right]^{-1} = 1 - J^{(1)}(p^2) - J^{(2)}(p^2)
+ \left[ J^{(1)}(p^2) \right]^2 + \ldots
\ee

According to eq.~(\ref{gh_se}), the ghost propagator is 
proportional to
\be
G(p^2) = 1 + G^{(1)}(p^2) + G^{(2)}(p^2) + \ldots
\ee
The ghost self energy (which is inverse to the propagator) 
is proportional to
\bea  
\label{Gcomments}
\left[ G(p^2) \right]^{-1} &=&
1 - G^{(1)}(p^2) - G^{(2)({\rm irred})}(p^2) + \ldots
\nn \\
&=& 1 - G^{(1)}(p^2) - G^{(2)}(p^2) + \left[ G^{(1)}(p^2) \right]^2
+ \ldots
\eea
Note that the one-loop contribution to the ghost self energy gives
$-G^{(1)}(p^2)$. Two-loop contributions to the ghost self
energy are shown in Fig.~4. They give $-G^{(2)({\rm irred})}(p^2)$. 
According to eq.~(\ref{Gcomments}), the two-loop contribution 
to the ghost propagator
consists of two parts, the irreducible one and the reducible one, 
\be
\label{G+G}
G^{(2)}(p^2) = G^{(2)({\rm irred})}(p^2) + G^{(2)({\rm red})}(p^2),
\ee
where $G^{(2)({\rm red})}(p^2)=\left[ G^{(1)}(p^2) \right]^2$.

One-loop results in arbitrary space-time dimension are available
e.g.\ in \cite{Muta,DOT1} (see also in Appendix~A).
When we expand them in $\ep$ and keep the terms up to
the order $\ep$, we get
\bea
J^{(1,\xi)}(p^2) = 
C_A \frac{g^2\; \eta}{(4\pi)^{n/2}} \; (-p^2)^{-\ep}
\left\{
\frac{1}{\ep}\left(- \frac{5}{3}- \frac{1}{2}\xi \right)
- \frac{31}{9} +\xi - \frac{1}{4}\xi^2
\hspace{20mm}
\right.
\nn \\
\left.
+ \ep \left(- \frac{188}{27}+2\xi - \frac{1}{2}\xi^2 \right)
\right\} + {\cal{O}}(\ep^2),
\eea
\be
J^{(1,q)}(p^2) =
T \; \frac{g^2\; \eta}{(4\pi)^{n/2}} \; (-p^2)^{-\ep}
\left\{
\frac{4}{3\ep}  + \frac{20}{9} + \frac{112}{27} \ep
\right\}+ {\cal{O}}(\ep^2),
\ee
\be
\label{G1exp}
G^{(1)}(p^2) =
C_A \; \frac{g^2\; \eta}{(4\pi)^{n/2}} \; (-p^2)^{-\ep}
\left\{
\frac{1}{\ep}\left( \frac{1}{2} +\frac{1}{4}\xi \right)
+ 1 + 2 \ep
\right\} + {\cal{O}}(\ep^2).
\ee

Calculating the sum of one-particle irreducible
two-loop diagrams contributing to the gluon polarization operator
(shown in Fig.~3), we have obtained the following unrenormalized 
results:
\bea
J^{(2,\xi)}(p^2) =
C_A^2 \frac{g^4 \; \eta^2}{(4\pi)^n}  (-p^2)^{-2\ep}
\left\{
\frac{1}{\ep^2}
\left(- \frac{25}{12}\!+\!\frac{5}{24}\xi\!+\!\frac{1}{4}\xi^2 \right)
\!+\!\frac{1}{\ep}
\left(- \frac{583}{72}\!+\! \frac{113}{144}\xi\!-\! \frac{19}{24}\xi^2
      \!+\!\frac{3}{8}\xi^3 \right)
\right.
\hspace*{-3.5mm}
\nn \\
\left.
- \frac{14311}{432}+ \zeta_3+ \frac{425}{864}\xi+ 2\xi \zeta_3
- \frac{71}{72}\xi^2+ \frac{9}{16}\xi^3 + \frac{1}{16} \xi^4
\right\}+{\cal{O}}(\ep),
\hspace{4mm}
\eea
\bea
J^{(2,q)}(p^2) =
C_A T \frac{g^4 \; \eta^2}{(4\pi)^n} \; (-p^2)^{-2\ep}
\left\{
\frac{1}{\ep^2}
\left(\frac{5}{3}- \frac{2}{3}\xi \right)
+\frac{1}{\ep}
\left(\frac{101}{18} + \frac{8}{9}\xi - \frac{2}{3}\xi^2 \right)
\right.
\nn \\
\left.
+ \frac{1961}{108}+ 8 \zeta_3+ \frac{142}{27}\xi- \frac{22}{9}\xi^2
\right\}
\nn \\
+ C_F T \frac{g^4 \; \eta^2}{(4\pi)^n} \; (-p^2)^{-2\ep}
 \left\{ \frac{2}{\ep}+ \frac{55}{3} - 16 \zeta_3 \right\}
+{\cal{O}}(\ep).
\hspace{20mm}
\eea

Calculating the sum of the contributions (Fig.~4) to the ghost
self energy (with a minus sign, cf. eq.~(\ref{Gcomments})), 
we obtain 
\bea
G^{(2,\xi)({\rm irred})}(p^2) =
C_A^2 \frac{g^4 \; \eta^2}{(4\pi)^n} \; (-p^2)^{-2\ep}
\left\{
\frac{1}{\ep^2}\left( 1+\frac{3}{16}\xi- \frac{3}{32}\xi^2\right)
+\frac{1}{\ep}\left(\frac{67}{16}- \frac{9}{32}\xi\right)
\right.
\nn \\
\left.
+ \frac{503}{32} - \frac{3}{4} \zeta_3- \frac{73}{64}\xi
+ \frac{3}{8}\xi^2 - \frac{3}{16}\xi^2 \zeta_3
\right\} + {\cal{O}}(\ep),
\eea
\be
G^{(2,q)}(p^2) =
C_A T \frac{g^4 \; \eta^2}{(4\pi)^n} \; (-p^2)^{-2\ep}
 \left\{ - \frac{1}{2\ep^2} - \frac{7}{4\ep} 
         - \frac{53}{8} \right\} + {\cal{O}}(\ep).
\ee
Note that there is no reducible part in $G^{(2,q)}$.
The reducible part of $G^{(2,\xi)}$ is given by the square
of eq.~(\ref{G1exp}), 
\be
G^{(2,\xi)({\rm red})}(p^2) 
= C_A^2 \frac{g^4 \; \eta^2}{(4\pi)^n} (-p^2)^{-2\ep}
\left\{ 
\frac{1}{\ep^2}
\left(\frac{1}{4}+\frac{1}{4}\xi+\frac{1}{16} \xi^2 \right)
+\frac{1}{\ep} \left( 1  + \frac{1}{2}\xi \right)
+ 3 + \xi
\right\} + {\cal{O}}(\ep).
\ee
Therefore, using eq.~(\ref{G+G}) we get
\bea
G^{(2,\xi)}(p^2) 
= C_A^2 \frac{g^4 \; \eta^2}{(4\pi)^n} (-p^2)^{-2\ep}
\left\{
\frac{1}{\ep^2}
\left( \frac{5}{4} +\frac{7}{16}\xi -\frac{1}{32}\xi^2 \right)
+\frac{1}{\ep}
\left( \frac{83}{16}+ \frac{7}{32}\xi \right)
\right.
\nn \\
\left.
+ \frac{599}{32}- \frac{3}{4} \zeta_3- \frac{9}{64}\xi
+ \frac{3}{8}\xi^2 - \frac{3}{16} \xi^2\zeta_3
\right\}+ {\cal{O}}(\ep).
\eea

\section{WST identity at the two-loop level}
\setcounter{equation}{0}

Due to the differential WST identity, we get the representations
(\ref{T1-WST}) and (\ref{T2-WST}) for the functions $T_i(p^2)$.
In the massless case, all one-loop expressions are 
proportional to $(p^2)^{-\ep}$,
whereas two-loop expressions contain $(p^2)^{-2\ep}$.
Thus, the differentiations in (\ref{T2-WST}) become trivial.
Expanding in $g^2$, we get\footnote{We take into account that 
(in the massless case) $G(0)=1$.}
\be
T_1^{(1)}(p^2) = a_3^{(1)}(p^2) + G^{(1)}(p^2) + J^{(1)}(p^2) ,
\ee
\bea
T_1^{(2)}(p^2) = a_3^{(1)}(p^2) 
\left[G^{(1)}(p^2) + J^{(1)}(p^2) \right]
+ G^{(1)}(p^2) J^{(1)}(p^2)
\nn \\
+ a_3^{(2)}(p^2) + G^{(2)}(p^2) + J^{(2)}(p^2) ,
\eea
\be
T_2^{(1)}(p^2) = 
2 T_1^{(1)}(p^2)
-2 \left[ (1-\ep) J^{(1)}(p^2) + (1+\ep) a_2^{(1)}(p^2) 
                  + {\widetilde{a}}_2^{(1)}(p^2) \right],
\ee
\bea
T_2^{(2)}(p^2) = 2 T_1^{(2)}(p^2)
-2 \left[ J^{(1)}(p^2) a_2^{(1)}(p^2)
+ J^{(1)}(p^2) \widetilde{a}_2^{(1)}(p^2) 
\right.
\nn \\
\left.
+ (1-2\ep) J^{(2)}(p^2) + (1+2\ep) a_2^{(2)}(p^2)
+ \widetilde{a}_2^{(2)}(p^2)
\right].
\eea

Substituting the expressions for ghost-gluon vertex and two-point
functions, we arrive at the same results as given in 
(\ref{T12xi})--(\ref{T22q}).

\section{Renormalization}
\setcounter{equation}{0}

To begin this section, we would like to explain 
why the zero-momentum limit of the three-gluon vertex, 
as well as the relevant limits of the ghost-gluon vertex, are
infrared finite, i.e. we do not get any $1/\ep$ poles of 
infrared (on-shell) origin. The main argument is just power counting.

Consider a triple vertex $V_0$ (part of a two-loop diagram)
to which are attached the zero-momentum external line, 
together with two adjacent propagators carrying
{\it the same} loop momentum $q$.
In the case of a scalar (say, $\phi^3$) theory, one would get
$1/(q^2)^2$ in the integrand, leading to an infrared divergency.
However, in QCD the vertex $V_0$ can be either 
(i) a three-gluon vertex, 
(ii) a ghost-gluon vertex, or 
(iii) a quark-gluon vertex.
Effectively, the power of the gluon or ghost propagator in QCD is
$1/(q^2)$,
whereas for the massless quark propagator we get $1/q$.
Therefore, the case (iii) is infrared finite, since we get
only $1/q^2$ from the two quark propagators (no $q$-dependent
factor from the vertex).
In the cases (i) and (ii), we get $1/(q^2)^2$ from the two 
gluon (or ghost) propagators.
However, we also get a momentum-dependent factor from the
three-gluon (or ghost-gluon) vertex $V_0$, which cannot contain 
any momentum other than $q$ (since the external momentum is
zero). This gives in the numerator a factor which is linear in $q$,
so that effectively the infrared behaviour is
just $1/q^3$, i.e. we have no infrared divergency.
When the zero-momentum line is attached to the four-gluon vertex
like e.g. in diagrams ($h$) and ($h'$) in Fig.~1, 
we may also get two propagators carrying the same momentum $q$.
However, a similar
power counting shows that there are no infrared singularities.
For example, in diagrams ($h$) and ($h'$) an extra momentum $q$  
appears in the numerator from the one-loop self-energy-type 
insertion. 
This explains why {\em all} singularities 
in this limit are of ultra\-violet origin, and therefore should be 
removed by renormalization. 

In this paper we adopt the modification of the
renormalization prescription by `t~Hooft \cite{Hooft}, 
corresponding to the so-called $\overline{\mbox{MS}}$ 
scheme \cite{MSbar}. 
In this section (and in Appendix~B), the notations 
$\xi$, $\alpha$, $g^2$, etc.\ (without subscript) correspond 
to the {\em renormalized} (in the $\overline{\mbox{MS}}$ scheme) 
quantities. 
In previous sections (and in Appendix~A), they should be understood
as the {\em bare} quantities $\xi_B$, $\alpha_B$, $g_B^2$, etc.

The renormalization
constants $Z_{\Gamma}$ relating the dimensionally-regularized
one-particle-irreducible Green functions to the renormalized 
ones,
\begin{equation}
\label{renormalization}
\Gamma^{{\rm (ren)}}\left(\left\{\frac{p_i^2}{\mu^2}\right\},
\alpha,g^2\right)=
\lim_{\varepsilon \to 0}
\left[
Z_{\Gamma}\left(\frac{1}{\varepsilon},\alpha,g^2\right)
\Gamma\left(\{p_i^2\},\alpha_B,g^2_B,\varepsilon\right)
\right],
\end{equation}
look in this scheme like
\begin{equation}
\label{Z_Gamma}
Z_{\Gamma}\left( \frac{1}{\varepsilon},\alpha,g^2
         \right)=1+\sum_{j=1}^{\infty}
C_{\Gamma}^{[j]}(\alpha,g^2)  \frac{1}{\varepsilon^j},
\end{equation}
where $\alpha=1-\xi$.
In eq.~(\ref{renormalization}) $\mu$ is the renormalization 
parameter with the dimension of mass. 
It is assumed that on the r.h.s.
of eq.~(\ref{renormalization}) the squared bare charge
$g_B^2$ and the bare gauge parameter $\alpha_B$
must be substituted in terms of renormalized ones,
multiplied by appropriate $Z$ factors (cf. eqs.~(\ref{g_Z})
and (\ref{alpha_Z})).

We use the following definitions for renormalization factors:
\begin{eqnarray}
\label{defZ1}
&&\Gamma_{\mu_1 \mu_2 \mu_3 }^{{\rm (ren)}}(p_1,p_2,p_3)
 =Z_1\;\Gamma_{\mu_1 \mu_2 \mu_3 }(p_1,p_2,p_3), 
\\
\label{defZ1tilde}
&& 
\Pi^{{\rm (ren)}\;a_1 a_2}_{\mu_1 \mu_2}(p)
   =Z_3 \;\Pi^{a_1 a_2}_{\mu_1 \mu_2}(p), 
\\
\label{defZ3}
&&\widetilde{\Gamma}_{\mu}^{{\rm (ren)}\; a_1 a_2 a_3}(p_1,p_2,p_3)
=\widetilde{Z}_1 \; \tilde{\Gamma}_{\mu}^{a_1 a_2 a_3}(p_1,p_2,p_3), 
\\
\label{defZ3tilde}
&&\widetilde{\Pi}^{{\rm (ren)}\; a_1 a_2}(p^2)
 =\widetilde{Z}_3 \; \widetilde{\Pi}^{a_1 a_2}(p^2),
\end{eqnarray}
where $\Pi^{a_1 a_2}_{\mu_1 \mu_2}(p)$ and 
$\widetilde{\Pi}^{a_1 a_2}(p^2)$ are the gluon polarization
operator and the ghost self energy, respectively.
For the scalar amplitudes, eqs.~(\ref{defZ3})--(\ref{defZ3tilde})
mean that $J(p^2)$ and $G(p^2)$ should be renormalized by means 
of $Z_3$ and $\widetilde{Z}_3^{-1}$, respectively.
Furthermore, according to eqs.~(\ref{defZ1})--(\ref{defZ1tilde}) 
the three-gluon amplitudes ($T_1$ and $T_2$)
should be renormalized using $Z_1$, whereas for the ghost-gluon 
functions ($a_3,\; e_3,\; a_2$ and $e'_2$) one should use 
$\widetilde{Z}_1$.

The WST identity requires that
\be
\label{WST-Z}
\frac{Z_3}{Z_1}=\frac{\widetilde{Z}_3}{\widetilde{Z}_1} .
\ee
If this condition is satisfied, the WST identity is valid 
for the renormalized quantities, too. 

Using (\ref{WST-Z}), the bare coupling constant $g_B^2$ 
can be chosen (in the $\overline{\mbox{MS}}$ scheme) 
as\footnote{The factor $(e^{\gamma}/(4\pi))^{\ep}
= \exp{[\varepsilon(\gamma-\ln(4\pi))]}$ in eq.~(\ref{g_Z})
represents the difference between the $\overline{\mbox{MS}}$ 
and MS schemes (cf.\ also eq.~(\ref{eta})).}
\begin{equation}
\label{g_Z}
g_B^2=
\left(\frac{\mu^2 e^{\gamma}}{4\pi}\right)^{\ep} 
g^2 \widetilde{Z}_1^2
Z_3^{-1}\widetilde{Z}_3^{-2}
=\left(\frac{\mu^2 e^{\gamma}}{4\pi}\right)^{\ep} 
g^2 Z_1^2 Z_3^{-3}.
\end{equation}
The gauge parameter $\alpha=1-\xi$ is renormalized as 
\be
\label{alpha_Z}
\alpha_B=Z_3 \alpha,
\hspace{10mm} \mbox{so that} \hspace{10mm}
\xi_B = 1- Z_3 (1-\xi) .
\ee

Below we shall use the following notation:
\be
h \equiv \frac{g^2}{(4\pi)^2} = \frac{\alpha_s}{4\pi} , 
\hspace{10mm} 
\mbox{where} 
\hspace{10mm}
\alpha_s\equiv \frac{g^2}{4\pi} .
\ee

The two-loop-order results for the
renormalization factors have been obtained in \cite{Jones,VlaTar,EgTar} 
(see also in ref.~\cite{PT-QCD}). 
For completeness, we list the corresponding expressions
in Appendix~B. 

Using eqs.~(\ref{T1xi})--(\ref{T2q}), (\ref{T12xi})--(\ref{T22q}),
(\ref{defZ1}) and (\ref{Z1}), 
we obtain the renormalized 
scalar amplitudes appearing in the three-gluon vertex
(cf. eq.~(\ref{BL-decomp})),
\bea
\label{T1ren}
T_1^{{\rm (ren)}} =1
 +h \left[
 C_A\left(-\frac{35}{18}+\frac{1}{2}\xi-\frac{1}{4}\xi^2\right)
 + \frac{20}{9} T
    \right]
\hspace{50mm}
\nn \\
+h^2 \left[
 C_A^2  \left(
 - \frac{4021}{288} - \frac{1}{4} \zeta_3
 -\frac{2317}{576} \xi  + \frac{15}{8}\xi \zeta_3 
 + \frac{113}{144} \xi^2  -\frac{1}{16} \xi^3 
 + \frac{1}{16} \xi^4  \right)
\right. 
\nn \\
\left. 
 + C_A T  \left( \frac{875}{72}  + 8 \zeta_3 + \frac{20}{9}\xi
       -\frac{10}{9} \xi^2 \right)
+ C_F T \left( \frac{55}{3} - 16 \zeta_3 \right)  
\right] + {\cal{O}}(h^3), 
\eea
\bea
\label{T2ren}
T_2^{{\rm (ren)}}=
h\left[
C_A \left(- \frac{4}{3} - 2\xi + \frac{1}{4} \xi^2 \right) 
+ \frac{8}{3} T \right] 
+h^2 \left[ C_A T \left(
 \frac{157}{9}-\frac{37}{18}\xi-\frac{2}{9} \xi^2 \right)
+8 C_F T
 \right. 
\nn \\
\left. 
+ C_A^2 \left(
-\frac{641}{36}- \zeta_3
+ \frac{5}{18} \xi-\frac{1}{2} \xi \zeta_3 
-\frac{287}{144} \xi^2 + \frac{19}{16} \xi^3
 - \frac{1}{8} \xi^4 
\right) \right] 
+ {\cal{O}}(h^3).  
\eea
Here and henceforth, 
we put $p^2=-\mu^2$ in the renormalized expressions.
In Feynman gauge ($\xi=0$), our expressions agree with 
eq.~(B4) from \cite{BL}. However, the one-loop part of the result
for $T_2$ in an arbitrary (non-Feynman) gauge disagrees with 
eq.~(A10) from
\cite{BL}\footnote{Cf. footnote~19 on p.~4101 of \cite{DOT1}. 
In {\em our} notation, in the $hC_A$ part of (\ref{T2ren}) the term
$\frac{1}{4}\xi^2$ is missing in \cite{BL}.}.

The renormalized expressions for two-point functions are
\bea
\label{Jren}
J^{{\rm (ren)}}=1+h\left[ 
C_A \left( -\frac{31}{9} + \xi - \frac{1}{4}\xi^2 \right)
+ \frac{20}{9} T \right]
\hspace{60mm}
\nn \\
+ h^2 \left[
C_A^2 \left( -\frac{3245}{144}+\zeta_3-\frac{287}{96}\xi+2\xi\zeta_3
             +\frac{61}{72}\xi^2-\frac{3}{16}\xi^3+\frac{1}{16}\xi^4
      \right)
\right.
\hspace{12mm}
\nn \\
\left.
+ C_A T \left( \frac{451}{36} + 8\zeta_3 + \frac{10}{3}\xi
              -\frac{10}{9}\xi^2 \right)
+ C_F T \left( \frac{55}{3} - 16\zeta_3 \right) \right]
+ {\cal{O}}(h^3),
\eea
\be
G^{{\rm (ren)}}=1+h C_A + h^2 \left[ 
C_A^2 \left( \frac{997}{96} - \frac{3}{4}\zeta_3 - \frac{41}{64}\xi
             +\frac{3}{8}\xi^2 - \frac{3}{16} \xi^2 \zeta_3 \right)
- \frac{95}{24} C_A T \right]
+ {\cal{O}}(h^3).
\ee
In Feynman gauge, eq.~(\ref{Jren}) gives the same as the first
of eqs.~(B3) in ref.~\cite{BL}. 
Taking into account that
\be
\left[G^{-1}\right]^{{\rm (ren)}}
= 2 - G^{{\rm (ren)}}+h^2 C_A^2 + {\cal{O}}(h^3) ,
\ee
we have also confirmed the second of eqs.~(B3) in \cite{BL},
i.e.\ the result for the ghost self energy in Feynman gauge.

The renormalized expressions for the scalar functions
occurring in the ghost-gluon vertex are
\bea
a_3^{{\rm (ren)}}
=1 + \frac{1}{2} \; h \; C_A \; (1-\xi)
\hspace{98mm}
\nn \\
+ h^2 \left[ C_A^2 
\left( \frac{137}{48} - \frac{1}{2} \zeta_3 - \frac{299}{96}\xi
      -\frac{1}{8}\xi\zeta_3 + \frac{7}{16}\xi^2 
      + \frac{3}{16}\xi^2\zeta_3 \right) + \frac{1}{4} C_A T \right]
+ {\cal{O}}(h^3),
\eea
\be
p^2 e_3^{{\rm (ren)}} = \frac{1}{4} \; h \; C_A \; (2+\xi) 
+h^2 \left[ C_A^2
\left( \frac{20}{3} + \frac{1}{8}\zeta_3 - \frac{5}{12}\xi
+\frac{5}{16} \xi \zeta_3 - \frac{3}{16} \xi^2 \right)
- \frac{8}{3} C_A T \right] 
+ {\cal{O}}(h^3),
\ee
\bea
a_2^{{\rm (ren)}}
=1+ \frac{1}{4} \; h \; C_A \; \xi (1-\xi)
\hspace{97mm}
\nn \\
+ h^2 \; (1-\xi) \; \left[ C_A^2 
\left( \frac{167}{72} - \zeta_3 - \frac{43}{144}\xi - \frac{1}{16}\xi^2
       - \frac{1}{16}\xi^3 \right) 
- \frac{5}{9} C_A T (1-\xi) \right]
+ {\cal{O}}(h^3),
\eea
\bea
p^2 {e'}_2^{{\rm (ren)}}=
 \frac{1}{4} \; h \; C_A \; (1-\xi)(2-\xi)
\hspace{91mm}
\nn \\
+ h^2 (1\!-\!\xi) \left[ C_A^2 
\left( \frac{29}{36} \!+\! \frac{5}{8} \zeta_3 \!-\! \frac{5}{36} \xi
\!-\! \frac{3}{16}\xi\zeta_3 \!+\! \frac{1}{16} \xi^2 
+ \frac{1}{16} \xi^3 \right)
\!+\! \frac{5}{9} C_A T (1\!-\!\xi) \right] 
+ {\cal{O}}(h^3) .
\hspace{3.5mm}
\eea
We note that these functions are in the following correspondence
with the functions $G_{1,2}(p^2)$ used in \cite{BL}, eq.~(A3):
\be
a_3+p^2 e_3 \leftrightarrow 1+G_2 ,
\hspace{15mm}
a_2+p^2 e'_2 \leftrightarrow 1+G_1.
\ee
Using this connection, we have confirmed the two-loop-order
results for $G_1$ 
and $G_2$ in the Feynman gauge, eq.~(B5) of ref.~\cite{BL},
as well as the one-loop-order results for $G_1$
and $G_2$ in an arbitrary covariant gauge,
eq.~(A11) of \cite{BL}. 

\section{Conclusion}
\setcounter{equation}{0}

In the limit when one of the gluon momenta vanishes, we have
calculated the two-loop contributions to the three-gluon vertex,
in an arbitrary covariant gauge. In fact, we needed to calculate
two scalar functions, $T_1(p^2)$ and $T_2(p^2)$,
associated with different tensor structures, cf.\ eq.~(\ref{BL-decomp}).
Two independent ways of calculating these scalar functions
have been realized. One of them is 
based on the straightforward calculation of all diagrams
contributing to the two-loop three-gluon vertex shown in Fig.~1. 

Another way of determining $T_1(p^2)$ and $T_2(p^2)$ is based on 
exploiting the differential WST identity (\ref{WST-diff}). 
In this way, we obtain representations
of the scalar functions $T_1(p^2)$ and $T_2(p^2)$, 
eqs.~(\ref{T1-WST}) and (\ref{T2-WST}), in terms
of the functions occurring in the ghost-gluon vertex (Fig.~2),
its derivative (\ref{a_tilde_2}), the gluon polarization operator
(Fig.~3) and the ghost propagator (cf. Fig.~4). 
We have calculated all these
functions and confirmed the result of the straightforward
calculation.

The construction of the differential WST identity is of a
certain interest, since in this limit it {\it completely} defines
the three-gluon vertex, without leaving any ``undetected''
transverse contributions.

We have constructed renormalized expressions for all Green
functions involved.
Note that in the zero-momentum limit the three-gluon vertex has no
infrared (on-shell) singularities, this is a ``pure''
case for performing the ultraviolet renormalization.

The obtained results can be considered as the first step in 
constructing
expressions for the QCD vertices in more complicated cases,
including on-shell configurations and the general
off-shell case. In principle, the techniques for calculating 
the corresponding scalar integrals are already available
\cite{onshell,UD}.

\vspace{3mm}

{\bf Acknowledgements.}
We are grateful to S.A.~Larin for useful discussions.
O.~T. would like to thank Department of Physics, 
University of Bergen for warm hospitality during his
visit in 1996, when this work was started.
P.~O. would like to thank the DESY Theory group, where this work
was finished, for kind hospitality.
This research has been supported by the Research Council of Norway,
and by the Nordic project (NORDITA) 
`Fundamental constituents of matter'.

\appendix   
\section*{Appendix A: One-loop expressions for arbitrary $n$}
\setcounter{equation}{0}
\renewcommand{\thesection}{A}

At the zero-loop level, we have
\begin{equation}
a_3^{(0)}=a_2^{(0)}=1 , 
\; \; \; {\widetilde{a}}_2^{(0)} = 0 ,
\; \; \; d_2^{(0)}=0 , \; \; \;  J^{(0)}=G^{(0)}=1 ,
\end{equation}
and the r.h.s.\ of eq.~(\ref{WST-diff}) restores the zero-loop result
for the three-gluon vertex, 
\begin{equation}
\Gamma^{(0)}_{\mu_1 \mu_2 \mu_3}(p, -p, 0) =
2 g_{\mu_1 \mu_2} p_{\mu_3} - g_{\mu_1 \mu_3} p_{\mu_2}
         - g_{\mu_2 \mu_3} p_{\mu_1} .
\end{equation}

At the one-loop level, the expressions obtained in \cite{DOT1} give
the following results in the zero-momentum limit:
\begin{eqnarray}
a_3^{(1)}(p^2) &=&
\frac{g^2\;\eta}{(4\pi)^{n/2}} \;
\frac{C_A}{4} \; \kappa(p^2) \; (n-2) (1-\xi) ,
\\
p^2 e_3^{(1)}(p^2)
&=& -\frac{g^2\;\eta}{(4\pi)^{n/2}} \;
\frac{C_A}{8} \kappa(p^2) (n-4)
\left[ 2 + (n-3)\xi \right] ,
\\
a_2^{(1)}(p^2) &=&
\frac{g^2\;\eta}{(4\pi)^{n/2}} \;
\frac{C_A}{8} \; \kappa(p^2) \;  
(1- \xi) \left[ 4(n-3) - (n-4)\xi \right] ,
\\
p^2 d_2^{(1)}(p^2) &=&
\frac{g^2\;\eta}{(4\pi)^{n/2}} \;
\frac{C_A}{8} \; \kappa(p^2) \;
\left[ 2(n-6) - (5n-18)\xi + (n-4)\xi^2 \right] ,
\\
p^2 {e'}_2^{(1)}(p^2)
&=& -\frac{g^2\;\eta}{(4\pi)^{n/2}} \;
\frac{C_A}{8} \kappa(p^2)
(1-\xi) (2-\xi) (n-4) ,
\end{eqnarray}
\be
\widetilde{a}_2^{(1)}(p^2)  
= \frac{g^2\;\eta}{(4\pi)^{n/2}}  \frac{C_A}{32} \; \kappa(p^2)
\left\{ 8(n^2-6n+10) - 2 \xi (3n^2-26n+52)
+ \xi^2 (n-4) (n-6) \right\} .
\ee

In these equations,
\begin{equation}
\label{kappa}
\kappa(p^2) \equiv
- \frac{2}{(n-3) (n-4)} \; (-p^2)^{(n-4)/2}
= \frac{1}{\varepsilon (1-2\varepsilon)} \; (-p^2)^{-\varepsilon} .
\end{equation} 

The results for two-point functions are (cf. e.g.\ in 
\cite{Muta,DOT1}):
\bea
J^{(1)}(p^2) =
\frac{g^2\;\eta}{(4\pi)^{n/2}} \;
\frac{\kappa(p^2)}{(n-1)} \;
\left\{ - \frac{C_A}{8}\left[ 4(3n\!-\!2)+4(n\!-\!1)(2n\!-\!7)\xi
-(n\!-\!1)(n\!-\!4)\xi^2 \right]
\right.
\nonumber \\
\left. \frac{}{}
+ 2 T (n-2) \right\},
\hspace{4mm}
\eea
\be
G^{(1)}(p^2) =
\frac{g^2\;\eta}{(4\pi)^{n/2}} \;
\frac{C_A}{4} \; \kappa(p^2) \; \left[ 2 + (n-3) \xi \right] .
\ee

Taking into account that
\be
\left[ (a_2-p^2 d_2) J \right]^{(1)} 
= a_2^{(1)} - p^2 d_2^{(1)} + J^{(1)} ,
\ee
\bea
\left[ 
\left( p^2 d_2 + \widetilde{a}_2 
- p^2 \frac{\mbox{d}a_2}{\mbox{d}p^2} \right) J 
+ p^2 a_2 \frac{\mbox{d}J}{\mbox{d}p^2} \right]^{(1)}
&=& p^2 d_2^{(1)} + \widetilde{a}_2^{(1)}
- p^2 \frac{\mbox{d}a_2^{(1)}}{\mbox{d}p^2}
+ p^2 \frac{\mbox{d}J^{(1)}}{\mbox{d}p^2}
\hspace*{16mm}
\nn \\
&=& p^2 d_2^{(1)} + \widetilde{a}_2^{(1)}
- \frac{n-4}{2} a_2^{(1)} + \frac{n-4}{2}J^{(1)} ,
\hspace*{4mm}
\eea
we have checked that eq.~(\ref{WST-diff}) is
satisfied at the one-loop level, for an arbitrary $n$.
Furthermore,
\begin{equation}
\label{ad_}
a_2^{(1)}(p^2) - p^2 d_2^{(1)}(p^2) =
a_3^{(1)}(p^2) + G^{(1)}(p^2) =
\frac{g^2\;\eta}{(4\pi)^{n/2}} \;
\frac{C_A}{4} \; \kappa(p^2) \; (n- \xi) .
\end{equation}
Therefore, eq.~(\ref{adG}) 
(which follows from eq.~(\ref{WST-ord}))
is satisfied at the one-loop level.

\appendix
\section*{Appendix B: Renormalization factors}
\setcounter{equation}{0}
\renewcommand{\thesection}{B}

The expressions for the relevant two-loop-order renormalization factors
have been presented in refs.~\cite{Jones,VlaTar,EgTar}
(cf. also in \cite{PT-QCD}). 
For completeness, we present
the corresponding expressions here\footnote{As in section~8, 
the renormalized quantities $\xi=1-\alpha$, $g^2$, etc.\
are understood.}: 
\bea
\label{Z1}
Z_1=1
+\frac{h}{\ep} \left[ C_A\left(\frac{2}{3}+\frac{3}{4}\xi\right)
                      -\frac{4}{3}T \right]
+h^2 \left\{
C_A T \left[ \frac{1}{\ep^2}\left(\frac{5}{2}-\xi\right)
-\frac{25}{12\ep} \right]
 - \frac{2}{\ep}C_F T
\right.
\hspace{10mm}
\nn \\
\left.
+C_A^2\left[ \frac{1}{\ep^2}
\left(-\frac{13}{8}-\frac{7}{16}\xi+\frac{15}{32}\xi^2\right)
+\frac{1}{\ep}
\left(\frac{71}{48}+\frac{45}{32}\xi-\frac{3}{16}\xi^2\right)
\right]
\right\} + {\cal{O}}(h^3),
\eea
\be
\label{Z1tilde}
\widetilde{Z}_1=1-\frac{h}{2\ep} C_A (1-\xi)
+h^2 C_A^2 (1-\xi)
\left[
\frac{1}{\ep^2} \left(\frac{5}{8} - \frac{1}{4}\xi \right)
+ \frac{1}{\ep} \left( -\frac{3}{8}+\frac{1}{16}\xi \right)
\right]
+ {\cal{O}}(h^3),
\ee
\bea
\label{Z3}
Z_3=1  
+\frac{h}{\ep}\left[ C_A \left(\frac{5}{3}+\frac{\xi}{2}\right)
-\frac{4}{3}T \right]
+h^2\left\{
C_A T \left[ \frac{1}{\ep^2}
\left( \frac{5}{3} - \frac{2}{3}\xi \right) - \frac{5}{2\ep} \right]
- \frac{2}{\ep} C_F T
\right.
\hspace{10mm}
\nn \\
\left.
+C_A^2 \left[ \frac{1}{\ep^2}
\left( -\frac{25}{12} + \frac{5}{24}\xi + \frac{1}{4}\xi^2 \right)
+\frac{1}{\ep}
\left( \frac{23}{8} + \frac{15}{16}\xi - \frac{1}{8}\xi^2 \right)
\right] \right\}
+ {\cal{O}}(h^3),
\eea   
\bea
\label{Z3tilde}
\widetilde{Z}_3=1+\frac{h}{\ep} C_A
\left( \frac{1}{2} + \frac{1}{4}\xi \right)
+ h^2 \left\{
C_A^2 \left[
\frac{1}{\ep^2}
\left( -1 - \frac{3}{16}\xi + \frac{3}{32}\xi^2 \right)
+ \frac{1}{\ep}\left(\frac{49}{48}-\frac{1}{32}\xi\right)
\right]
\right.
\nn \\ 
\left.
+ C_A T \left( \frac{1}{2\ep^2} - \frac{5}{12\ep} \right)
\right\}
+ {\cal{O}}(h^3),
\eea
where $\ep=(4-n)/2$ and $h=g^2/(4\pi)^2$.
One can check that eqs.~(\ref{Z1})--(\ref{Z3tilde})
obey the WST identity (\ref{WST-Z}), so only three of them
are independent. Using the results for
unrenormalized Green functions, we have performed
an independent check on these $Z$ 
factors\footnote{Note that the two-loop results for $Z$ factors 
in the MS scheme are of the same form as in the
${\overline{\mbox{MS}}}$ scheme; the only difference
is that $g^2$ in the definition of $h$ should be understood
as the renormalized squared charge in the MS scheme.}.

The results for these renormalization factors (without fermionic
contributions, i.e. for the pure Yang--Mills theory) were
first presented in \cite{Jones} (Feynman gauge) and \cite{VlaTar}
(an arbitrary covariant gauge). 
The complete results in an arbitrary covariant gauge, including the
fermionic contributions, were presented in \cite{EgTar}
(cf. also in \cite{PT-QCD}).
In \cite{EgTar}, the renormalization factors $Z_3$ and
$\widetilde{Z}_3$ were denoted as $Z_2$ and $\widetilde{Z}_2$.
There was an obvious misprint in the last term of the expression for
$Z_2$ where $\frac{\alpha^2}{2}T^2$ should read
$\frac{C_2}{2}tN$ (in their notation, $T^2\leftrightarrow C_F$,
$C_2\leftrightarrow C_A$, $tN \leftrightarrow T$).
We note that this misprint was copied over to 
the review \cite{Narison} and the textbook \cite{Muta}.
In \cite{Muta}, in the end of the first line of eq.~(C.6) 
for $\widetilde{Z}_3$, the term $\alpha_R^2 C_F$ should read
$C_G T_R N_f$ ($\alpha_R$ is the renormalized gauge parameter,
$C_G\leftrightarrow C_A$). Then, in the beginning of the last line
of eq.~(C.5) for $Z_3$,
$\frac{1}{8}C_G$ should read $\frac{1}{8}C_G^2$.
There are several misprints in eq.~(2.30b) of \cite{Narison}.
The term 
$\frac{\alpha_G^2}{2}\left(\frac{1}{4}\right)\frac{N^2-1}{2N}$ 
should read
$\frac{N}{2}\left(\frac{1}{4}\right)\frac{n}{2}$
($\alpha_G$ is the renormalized gauge parameter,
$n\leftrightarrow N_f$, $\frac{n}{2}\leftrightarrow T$).
In the previous term, $\frac{N}{4}$ should read $\frac{N^2}{4}$.
In the term involving $\frac{5}{12}$, the ``factor'' 
$\frac{n}{8}$ with the following bracket should be removed.
In the one-loop-order part, $\frac{\alpha_G}{3}$ should read
$\frac{\alpha_G}{2}$, cf. eq.~(2.30a).
Finally, in eq.~(2.31b) for $\widetilde{Z}_1$, the
one-loop-order contribution should be multiplied
by $\frac{1}{4}$, cf. eq.~(2.31a). 

Using the $1/\ep$ term of the renormalization factor $Z_{\Gamma}$
(cf. eq.~(\ref{Z_Gamma})), one can obtain the corresponding
anomalous dimension $\gamma_{\Gamma}$ via 
\be
\label{an_dim}
\gamma_{\Gamma}\left(\alpha,g^2\right)=
g^2 \frac{\partial}{\partial g^2} 
C_{\Gamma}^{[1]}\left(\alpha,g^2\right) .
\ee
We have checked that in the Feynman gauge $\xi=0$ ($\alpha=1$)
the results for the anomalous dimensions $\widetilde{\gamma}_1$,
$\gamma_3$ and $\widetilde{\gamma}_3$ coincide 
(in the two-loop approximation) with those from \cite{TVZh}.
The anomalous dimension $\gamma_1$ 
is related to the others via
$\gamma_1-\gamma_3=\widetilde{\gamma}_1-\widetilde{\gamma}_3$
(this follows from the WST identity (\ref{WST-Z})
and the definition (\ref{an_dim})).
Moreover, since (cf. in \cite{TVZh})
\be
\beta(g^2)=g^2\left[ 2 \widetilde{\gamma}_1\left(\alpha,g^2\right)
- \gamma_3\left(\alpha,g^2\right) 
- 2 \widetilde{\gamma}_3\left(\alpha,g^2\right) 
\right] ,
\ee
we obtain the same result for the two-loop $\beta$ function as those
given in \cite{Caswell,Jones,VlaTar,EgTar}\footnote{We just note 
two obvious misprints in \cite{Jones}: (i) in eq.~(23),
$Bu^2$ should read $Bu^5$ and (ii) in eq.~(24) (one-loop-order
part of the $\beta$ function) $\frac{8}{3}T(R)$
should read $\frac{4}{3}T(R)$. In eq.~(4) of \cite{Caswell},
the lower-case $z$'s should be understood.},
namely
\be
\frac{1}{g^2}\beta\left(g^2\right)
= h \left[ -\frac{11}{3} C_A + \frac{4}{3}T \right]
+ h^2 \left[ -\frac{34}{3} C_A^2 + \frac{20}{3} C_A T 
+ 4 C_F T \right] + {\cal{O}}\left(h^3\right).
\ee 
Higher terms of the $\beta$ function are available in 
refs.~\cite{TVZh,LV,LvRV}.


\begin{figure}[htb]
\begin{center}
\setlength{\unitlength}{1cm}
\begin{picture}(16,18)
\put(1,0)
\mbox{\epsfysize=22cm\epsffile{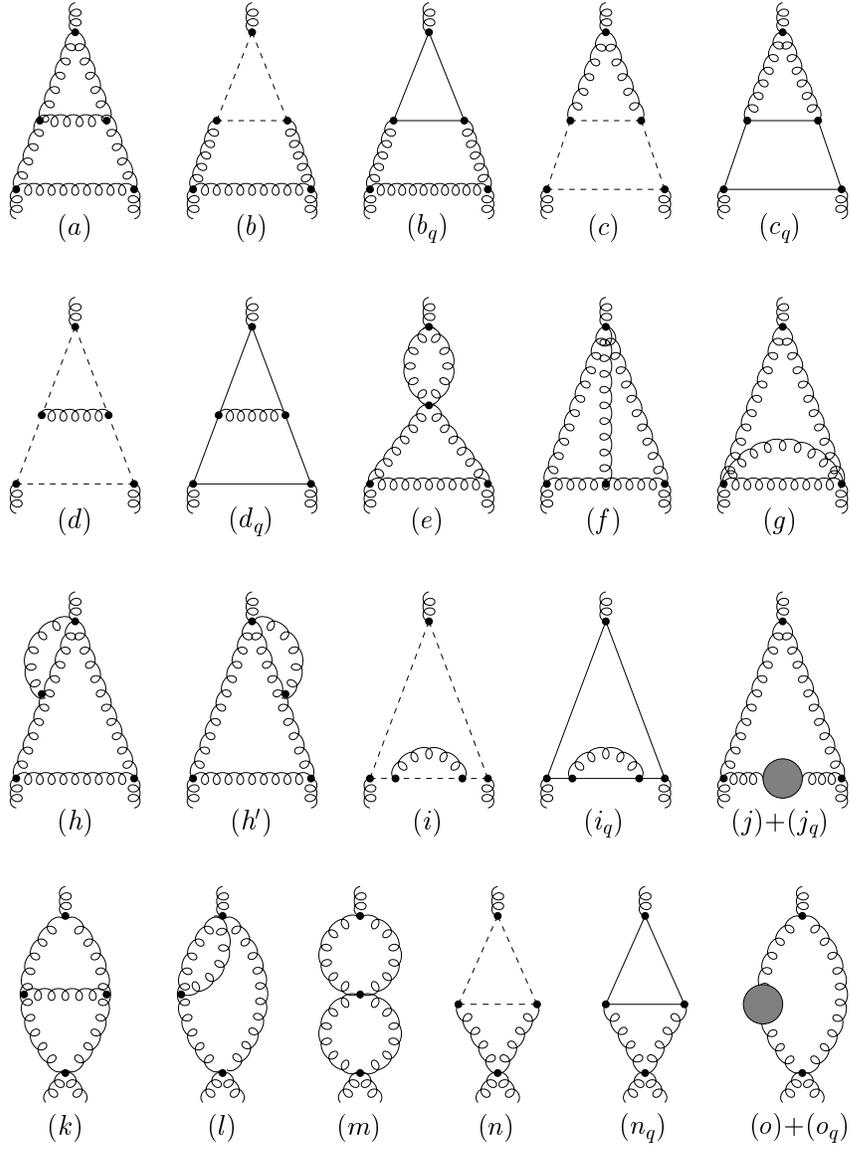}}
\end{picture}
\vspace*{-8mm}
\caption{Two-loop three-gluon vertex diagrams.}
\end{center}
\end{figure}
 
\begin{figure}[htb]
\begin{center}
\setlength{\unitlength}{1cm}
\begin{picture}(16,18)
\put(1,0)
\mbox{\epsfysize=22cm\epsffile{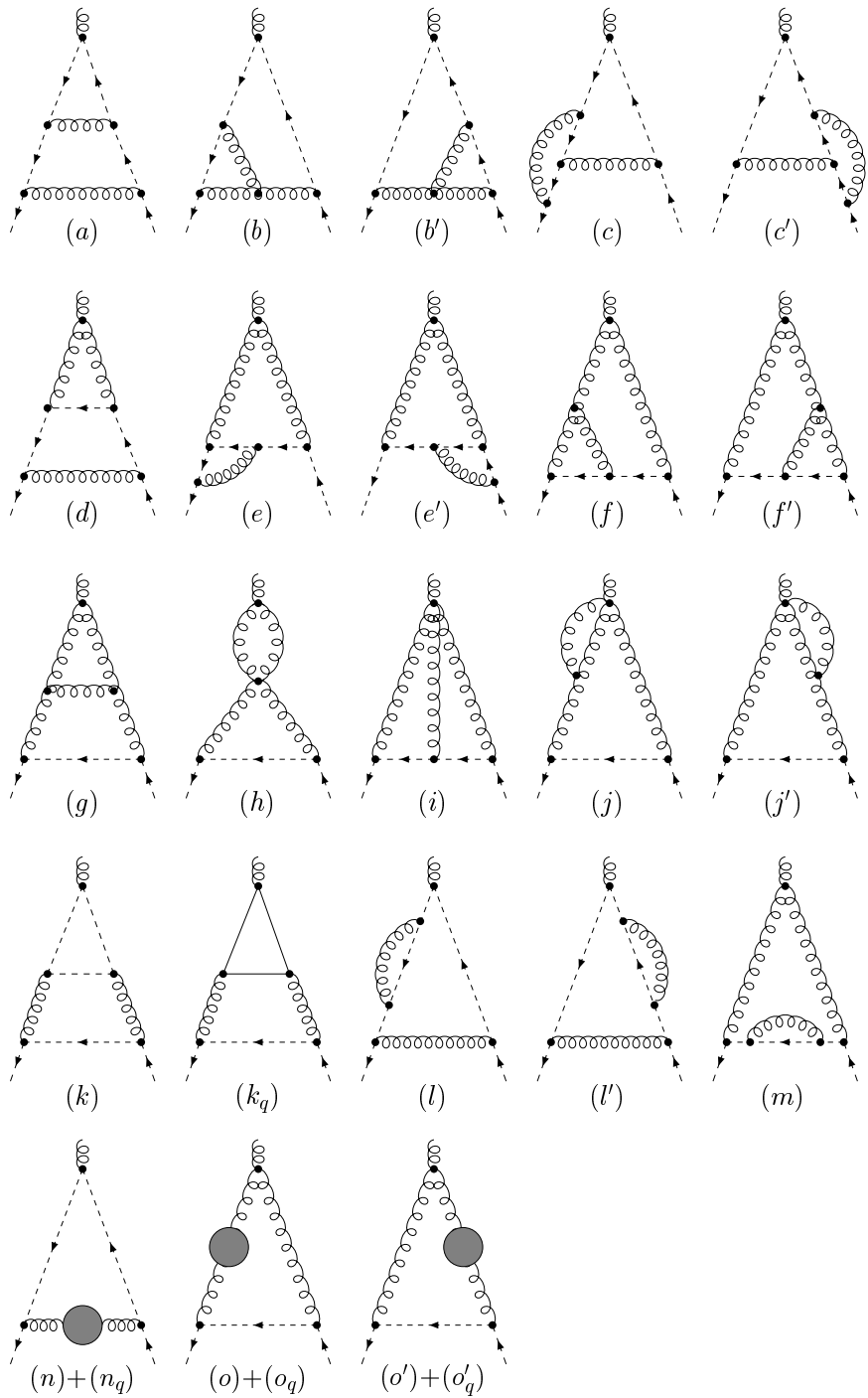}}
\end{picture}
\vspace*{-8mm}
\caption{Two-loop ghost-gluon vertex diagrams.}
\end{center}
\end{figure}
 
\begin{figure}[htb]
\begin{center}
\setlength{\unitlength}{1cm}
\begin{picture}(16,10)(1,-3)  
\put(1,0)
\mbox{\epsfysize=10cm\epsffile{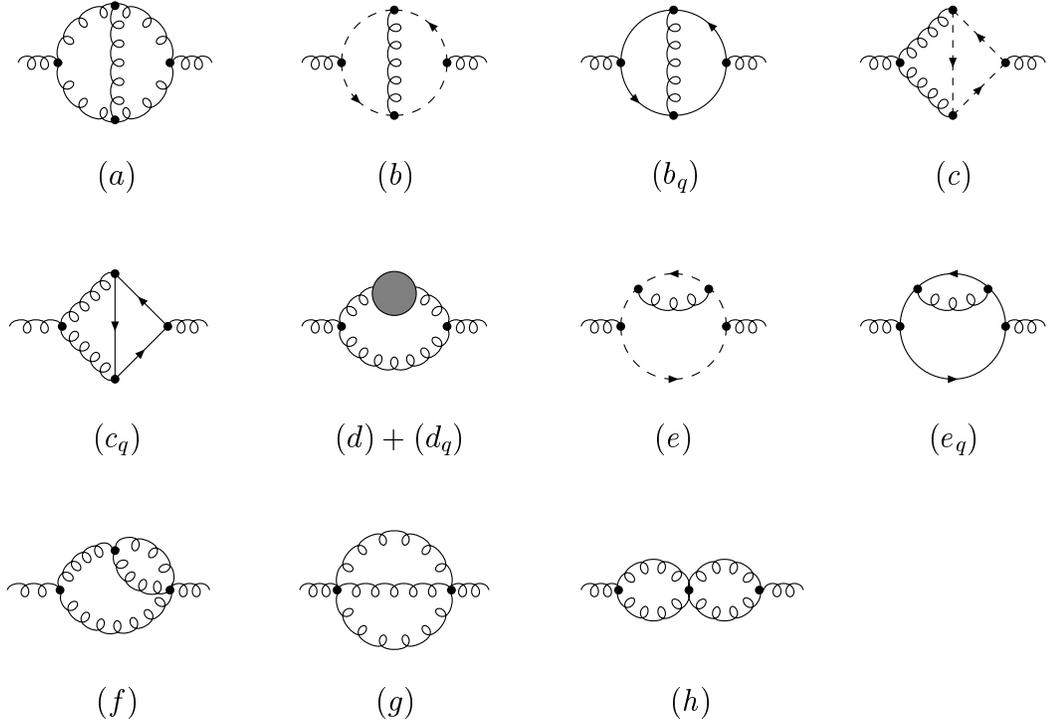}}
\end{picture}
\vspace*{-8mm}
\caption{Two-loop gluon polarization operator diagrams.}
\end{center}
\end{figure}
 
\begin{figure}[htb]
\begin{center}
\setlength{\unitlength}{1cm}
\begin{picture}(16,8)(1,-1)  
\put(1,0)
\mbox{\epsfysize=9cm\epsffile{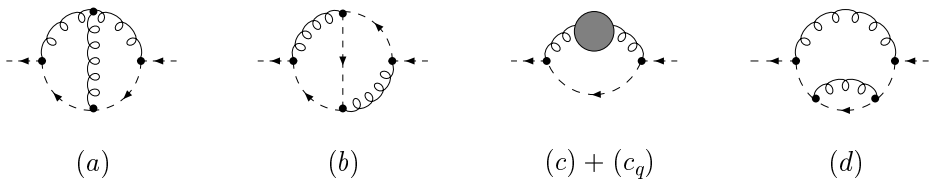}}
\end{picture}
\vspace*{-48mm}
\caption{Two-loop ghost self-energy diagrams.}
\end{center}
\end{figure}

\end{document}